\documentclass{article}
\usepackage{amsmath,graphicx,mlspconf}
\usepackage{xcolor}
\usepackage{amsfonts}
\usepackage{mathtools}
\usepackage{tikz}
\usepackage{hyperref}
\usepackage{stfloats}
\usepackage{booktabs} 
\usepackage{cite}
\toappear{2025 IEEE International Workshop on Machine Learning for Signal Processing, Aug.\ 31-- Sep.\ 3, 2025, Istanbul, Turkey}

\title{AUDIO PROTOTYPICAL NETWORK \\ FOR CONTROLLABLE MUSIC RECOMMENDATION}
\name{%
    \begin{tabular}{c}
    Fırat Öncel$^{1,6}$ \thanks{This research was enabled in part by compute resources provided by Mila (mila.quebec), and support provided by Calcul Québec and the Digital Research Alliance of Canada. We also acknowledge that this research was enabled in part by the Natural Sciences and Engineering Research Council of Canada (NSERC) and the NVIDIA Academic Grant Program for donating GPU hours used for this project. Laurent Charlin recognizes the support of the Canada CIFAR AI Chair Program, the Canada First Research Excellence Fund and IVADO.}%
    Emiliano Penaloza$^{2,6}$ Haolun Wu$^{3,6}$ Shubham Gupta$^{4,6}$ \\
    Mirco Ravanelli$^{1,6}$ Laurent Charlin$^{5,6}$ Cem Subakan$^{4,6}$
    \end{tabular}}
\address{%
    $^{1}$Concordia University $^{2}$Université de Montréal $^{3}$McGill University \\ $^{4}$Laval University $^{5}$HEC Montréal $^{6}$Mila - Quebec AI Institute}

\begin{document}
\maketitle

\begin{abstract}

Traditional recommendation systems represent user preferences in dense representations obtained through black-box encoder models. While these models often provide strong recommendation performance, they lack interpretability and controllability for users, leaving users unable to understand or control the system’s modeling of their preferences. This limitation is especially challenging in music recommendation, where user preferences are highly personal and often evolve based on nuanced qualities such as mood, genre, tempo, or instrumentation. 
In this paper, we propose an audio prototypical network for controllable music recommendation. This network expresses user preferences in terms of prototypes representative of semantically meaningful features pertaining to musical qualities. We show that the model obtains competitive recommendation performance compared to popular baseline models while also providing interpretable and controllable user profiles.
\end{abstract}


\usetikzlibrary{arrows}
\usetikzlibrary{quotes}
\usetikzlibrary{arrows.meta}
\def\BibTeX{{\rm B\kern-.05em{\sc i\kern-.025em b}\kern-.08em
    T\kern-.1667em\lower.7ex\hbox{E}\kern-.125emX}}

\tikzstyle{specialblock} = [draw, ultra thick, fill=blue!20, rectangle, 
    minimum height=3em, minimum width=4em]
\tikzstyle{block} = [draw, fill=lightgray, rectangle, 
    minimum height=3em, minimum width=4em]
\tikzstyle{sum} = [draw, fill=white, circle, node distance=1cm]
\tikzstyle{prod}   = [circle, minimum width=8pt, draw, inner sep=0pt, path picture={\draw (path picture bounding box.south east) -- (path picture bounding box.north west) (path picture bounding box.south west) -- (path picture bounding box.north east);}]
\tikzstyle{sumt}   = [circle, minimum width=8pt, draw, inner sep=0pt, path picture={\draw (path picture bounding box.east) -- (path picture bounding box.west) (path picture bounding box.south) -- (path picture bounding box.north);}]
\tikzstyle{input} = [coordinate]
\tikzstyle{output} = [coordinate]
\tikzstyle{pinstyle} = [pin edge={to-,thin,black}]
\tikzset{
tmp/.style  = {coordinate}, 
dot/.style = {circle, minimum size=#1,
              inner sep=0pt, outer sep=0pt},
dot/.default = 6pt 
}

\usetikzlibrary{arrows.meta}
\tikzstyle{dictsmall} = [draw, thick, fill=white!10, rectangle, 
    minimum height=1.0cm, minimum width=5cm] 
    \newcommand{\xshifts}{+4.7}

\section{Introduction}

Modern recommender systems often rely on techniques such as collaborative filtering methods, which represent users with dense vector embeddings that are difficult for their users to interpret and not meant to control recommendations. 

While previous work aims to improve the scrutability, being understandable and editable, of such systems by using keyword tags or natural language summaries to describe user preferences, such approaches are not universally applicable across domains. For instance, Siebrasse and Wald-Fuhrmann \cite{Siebrasse2023} demonstrate that using broad genres to describe someone’s musical taste can be misleading, as users with similar genre profiles may still have vastly different preferences. Their study shows that sub-genres, more closely tied to specific artists and musical elements, provide a more accurate representation of individual taste.

In light of these challenges, our work focuses on capturing user preferences through listenable audio clips, which transparently reflect the system's inferred understanding of their musical tastes. This encoding makes the system’s assumptions more interpretable and empowers users by allowing them to fully scrutinize and correct their profiles, offering control over how their preferences are represented and over their proposed recommendations. 

\tikzstyle{block} = [draw, fill=lightgray, rectangle, 
    minimum height=3em, minimum width=4em]

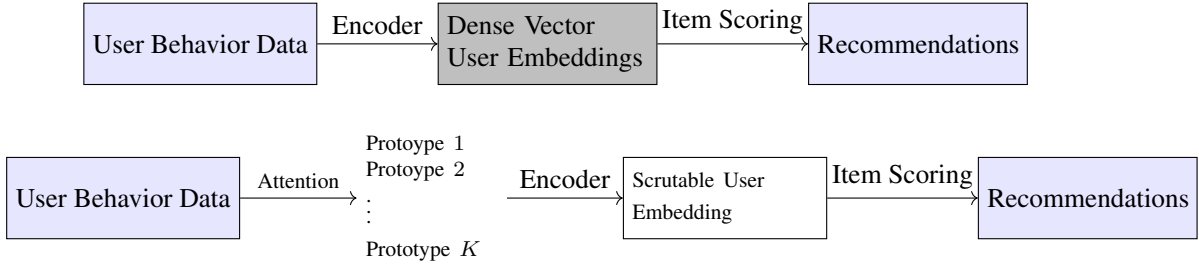
\begin{figure*}[t]
    \centering
    \resizebox{0.9\textwidth}{!}{
    \begin{tikzpicture}
        \node [block, fill=blue!10] (a) {User Behavior Data};
        \node [block, right of=a, xshift=3.5cm, text width=2.6cm] (b) {Dense Vector \\User Embeddings};
        \node [block, right of=b, fill=blue!10, xshift=3.8cm, text width=2.6cm] (c) {Recommendations};

        \node [block, below of=a, yshift=-1cm, xshift=-1cm, fill=blue!10] (d) {User Behavior Data};
        \node [right of=d, yshift=0cm, text width=1.7cm, xshift=3.0cm] (e) {\footnotesize{Protoype $1$\\ Protoype $2$ \\ $\vdots$ \\ Prototype $K$}};
        \node [block, fill=white, right of=e, xshift=2.8cm, text width=2.4cm] (f) {\footnotesize{Scrutable User Embedding}};
        \node [block, fill=blue!10, right of=f, xshift=3.7cm, text width=2.6cm] (g) {Recommendations};
        
        \draw [->] (a) -- node [anchor=south] {Encoder} (b);
        \draw [->] (b) -- node [anchor=south] {Item Scoring} (c);         

        \draw [->] (d) -- node [anchor=south] {\footnotesize Attention} (e);
        \draw [->] (e) -- node [anchor=south] {Encoder} (f);
        \draw [->] (f) -- node [anchor=south] {Item Scoring} (g);         
    \end{tikzpicture}
        }
    \caption{\textbf{(Top)} Classical Recommendation System Pipeline. (\textbf{Bottom}) The high-level pipeline for APRON. The User behavior Data (the audio features) are expressed in terms of listenable prototypes to create a scrutable user embedding. This embedding is then converted into recommendations.}
    \label{fig:highlevelapron}
\end{figure*}

We introduce APRON: \textbf{A}udio \textbf{PRO}totypical \textbf{N}etwork for music recommendation, where prototypes are listenable audio clips. We showcase the difference between a traditional recommendation system and APRON in Figure \ref{fig:highlevelapron}. APRON draws inspiration from prototypical networks (e.g.\ ProtoPNET \cite{protopnet, deformableprotopnet, willard2024looksbetterthatbetter, apnet, heinrich2024audioprotopnetinterpretabledeeplearning}), which are widely used in the Explainable AI (XAI) literature. 

APRON leverages an attention mechanism to create a weighted combination of prototype representations of users' historical interactions, ensuring an interpretable user representation.

Furthermore, by constraining the inferred prototype distribution to that of the recommended songs, we enable a fully steerable system, allowing users to scrutinize and adjust their profiles through simple modification of prototype weights. 

We demonstrate that our proposed methodology significantly enhances the controllability of the system's recommendations while maintaining performance comparable to fully black-box models. To evaluate controllability, we simulate user updates to their profiles, such as removing prototypes, and measure the differences in recommendations between the original and modified profiles.

We summarize our contributions as follows.
\begin{itemize}
    \item We propose APRON, a prototypical network for music that expresses the overall user preferences using prototypes composed of listenable audio clips.     

    \item We show that our model achieves a good controllability-accuracy tradeoff on the Million Song Dataset (MSD) \cite{bertin2011million, kim2023biasedjourneymsdaudiozip}. APRON obtains similar performance as several strong baselines while enabling users to control their recommendations directly by altering their learnt representation.

    \item To the best of our knowledge, this is the first work that allows users to scrutinize their recommendations using song-based prototypes, offering a offering a new interface for music recommendation and user interaction.
\end{itemize}

\subsection{Related Work}
Explainable recommendation has become an increasingly important topic as recommender systems grow more complex and opaque. Explainability has been approached post-hoc, using explanations based on dense features \cite{SLIM,vijayaraghavan2024robustexplainablerecommendation,zhang2020explainable} or more recently on LLM-produced explanations \cite{llmExp,luo2024unlockingpotentiallargelanguage}. Yet, as noted earlier, these explanations might not be \textit{actionable} by users or contain \textit{truthful} information \cite{huang2023surveyhallucinationlargelanguage}. Scrutable recommender systems present the user profile in a human-understandable and editable manner, enabling user interventions to directly influence the system's recommendations. This enhances actionability and truthfulness by allowing users to make meaningful changes that are transparently reflected in the system's behavior. Although they have many desirable properties, such systems have primarily been explored through the use of keywords or tags, which allow users to personalize their experience by selecting from a predefined collection \cite{tagsRecsys,llmExp,treeRecsys,scrutableRecsys, leszczynski2023talkwalksyntheticdata}. Representing a user's taste profile in this way can be limiting, as users may have to parse through an excessive collection of tags if they want to effectively customize their experience. Scrutable systems have shifted towards using natural language summaries to represent users, offering an alternative to keyword-based personalization \cite{positionScrutable,ramos-etal-2024-transparent,tears}. Instead of relying on keywords/tags, these systems generate a personalized summary using natural text. While this approach works well for domains suited to textual descriptions—such as movies, TV shows, or restaurants—it may not translate as effectively to other domains, like music or fashion, where user preferences might be difficult to express easily through text and could be better expressed through other mediums, such as audio or images. This highlights the need for more flexible approaches that can adapt scrutability to wider content types.
In this work, we address both limitations by enabling prototypes to attend to items in the user history, allowing us to maintain scrutability while offering a more personalized experience.
\section{Methodology}

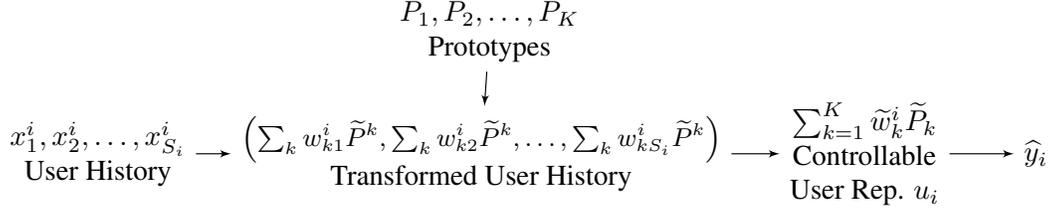
\begin{figure*}[t]
    \centering
    \resizebox{0.80\textwidth}{!}{
    \begin{tikzpicture}[auto, node distance=1.5cm,>=latex']
        \node [fill=none, xshift=0.7cm, align=center] (input) 
            {$x_1^i, x_2^i, \dots, x_{S_i}^i$\\User History};

    \node [right of=input, align=center, xshift=3.2cm] (ph) 
      {\small $\left(\sum_{k} w^i_{k1} \widetilde{P}^{k}, \sum_{k} w^i_{k2} \widetilde{P}^{k}, \dots, \sum_{k} w^i_{kS_i} \widetilde{P}^{k}\right)$ \\ Transformed User History};

        \node [above of=ph, xshift=0.1cm, align=center] (p) 
            {$P_1, P_2, \dots, P_K$\\Prototypes};

        \node [right of=ph, xshift=3.2cm, align=center] (u) 
            {$\sum_{k=1}^K \widetilde w^i_{k} \widetilde P_k$\\Controllable\\User Rep. $u_i$};

        \node [right of=u, xshift=0.6cm, align=center] (y) 
            {$\widehat y_i$};

        \draw [->] (input) -- (ph); 
        \draw [->] (p) -- (ph); 
        \draw [->] (ph) -- (u); 
        \draw [->] (u) -- (y); 
    \end{tikzpicture}
    }
    \caption{The pipeline of our model APRON. First, the user history is transformed by calculating the attention weights $w^i_{kj}$ using the prototypes $P_1, \dots, P_K$. Then the user representation $u_i$ is obtained by summing over the songs. Finally, the model output $\widehat y_i$ is computed by passing $u_i$ through a feedforward network $f(\cdot)$.}
    \label{fig:pipeline}
    \vspace{-0.5cm}
\end{figure*}

Our main goal is to express a user's historical interactions in terms of listenable prototypes, each associated with distinct musical concepts. In our experiments, musical concepts are encoded with tags corresponding to musical qualities (e.g. era, instrumentation, mood). Let us denote the user history for $i$'th user as
\begin{align}
\mathcal X_i = \{ x_1^i, x_2^i, \dots, x_{S_i}^i \},
\end{align}
where $S_i$ and $x_j^i \in \mathbb R^D$ respectively denote the total number of songs listened by user $i$, and the $D$-dimensional encoding of the $j$'th song listened by the $i$'th user. A reasonable way to construct the profile $u_i$ for user $i$ is by summing the representations of songs the user has listened to in the past,
\begin{align}
    u_i = \sum_{j=1}^{S_i} x_j^i.
\end{align}
 Such representations could then be processed by an encoder which directly provides recommendations. 
However, to impose a controllability constraint on the user profile, we constrain each song representation in terms of prototypes $\{P_1, \dots, P_K\}$, such that:
\begin{align}
    x_j^i = \sum_{k=1}^K w_{kj}^i P_k, \label{eq:userprofile}
\end{align}
where $P_k \in \mathbb R^D$ is the prototype that corresponds to the $k$'th musical tag. Each tag corresponds to a musical concept (e.g. indie rock, jazz, 90s, country, instrumental, more generally tags correspond to musical qualities). Note that each song can have more than a single tag (e.g. an instrumental song with two associated genres such as country and ballad). The weights $w_{kj}^i$ are parametrized using an attention layer, 
\begin{align}
    w_{kj}^i = \frac{\exp \left ( (P_k A_k^p)^\top (x_j^i A_j^x) \right )}{ \sum_{k'}^K \exp \left( (P_{k'} A_{k'}^p)^\top (x_j^i A_j^x) \right )} \label{eq:attention},     
\end{align}
where $A_k^p \in \mathbb R^{D\times D'}$, $A_j^x \in \mathbb R^{D \times D'}$ are learnable parameter matrices. Each user profile is then modelled as 
\begin{align}
    u_i = \sum_{k=1}^K \underbrace{ \sum_{j=1}^{S_i}  \frac{\exp \left ( (P_k A_k^p)^\top (x_j^i A_j^x) \right )}{ \sum_{k'}^K \exp \left( (P_{k'} A_{k'}^p)^\top (x_j^i A_j^x) \right )}}_{:=\widetilde w^i_{k}} \underbrace{(P_k A_k^p)}_{:=\widetilde P_{k}}
\end{align}

Note that unlike the Eq. \ref{eq:userprofile} the prototypes $P_k$ are transformed as well, such that we use the result of the vector-matrix product $P_k A_k^p$ as the Value vector in the attention calculation (similar to the standard query-key-value attention formulation). One difference from the standard query-key-value attention formulation is that we use the same learnable matrix for the key and value, since we observed that it 
results in a more controllable model. 

The output distribution over song recommendations $y_i \in \mathcal S^L$ (where $L$ is the number of songs in the catalog, and $\mathcal S^L$ denotes an $L$ dimensional probability simplex) is computed by obtaining the interpretable user profile from Eq.~\ref{eq:userprofile}, through a series of feed-forward layers denoted with $f(\cdot)$ followed by a softmax activation as:
\begin{align}
    \widehat y_i = a(f(u_i)), 
\end{align}
where $a(\cdot)$ is an activation function such as the Softmax or the Sigmoid. We describe the overall pipeline in Figure \ref{fig:pipeline}.

\section{Training Objectives}

For training the full system, we now demonstrate the three objectives employed as below.

\noindent \textbf{Recommendation Objective.} 
We train this system with a recommendation system loss that aims to minimize the divergence: 
\begin{align}
\mathcal L_\text{RecSys} = d(y \;\|\; \widehat y_i). 
\end{align}
The divergence is typically chosen as negative binary cross-entropy loss.

\noindent \textbf{Controllability Objective.} In addition to the recommendation system loss, to allow the system to be controllable, we construct a loss objective that minimizes the divergence between the aggregate prototype weights $\widetilde w^i_k$ and tag distribution that corresponds to the model output. We express this controllability loss $\mathcal L_{\text{controllability}}$ as follows: 
\begin{align}
    \mathcal L_{\text{controllability}} = d( \widetilde w^i \;\|\; T(\widehat y_i)),
\end{align}
where $T(\cdot)$ is a counting function that obtains the tag distribution given the songs selected with $\widehat y_i$. This loss imposes the constraint that, for user $i$, the tag distribution $T(\widehat y_i)$ that corresponds to the recommendation output of the model $\widehat y_i$ is as close as possible to the user's distribution over tag prototypes $\widetilde w^i \in \mathbb R^K$. For the choice of the divergence metric, we emprically observe that the Hellinger distance gives the best performance. Therefore in our experiments this is what we use for the controllability loss: 
\begin{align}
    \mathcal L_{\text{controllability}} = \sum_{i=1}^N \frac{1}{\sqrt{2}} \sqrt{\sum_{k=1}^K \left (\sqrt{\widetilde w^i_k} - \sqrt{T(\widehat y_i)_k} \right )^2},
\end{align}

\noindent \textbf{Prototype-separability Objective.}
We include a prototype-separability loss to make the prototypes as representative and distinct as possible of the associated music tags. For this, we enforce the transformed prototypes $\widetilde P_k$ to be classified as the associated tag, after passing these vectors through a linear layer $\phi(\cdot): D' \to K$. The corresponding loss is as follows: 
\begin{align}
    \mathcal L_{\text{prototype-sep}} = d( \phi(\widehat P_k)  \parallel  e_k),
\end{align}
where $e_k$ is the unit-vector that corresponds to the $k$'th tag, and for $d(.\|.)$ we used the standard cross-entropy loss for multi-way classification. We observed that this loss helps in avoiding solutions where the transformed prototypes collapse to very similar vectors. 

Finally, the overall training objective $\mathcal L$ is defined as a weighted sum of the above three objectives as follows, with relative strengths $\lambda_1$, $\lambda_2$. 
\begin{align}
    \mathcal L = \mathcal L_\text{RecSys}  + \lambda_1 \mathcal L_\text{prototype-sep} + \lambda_2 \mathcal L_\text{controllability}. 
\end{align}

\section{Experiments}
In this section, we evaluate the recommendation system performance of {APRON} along with other baseline models applicable for music recommendation. We also provide experimental results for controllability analysis of {APRON}. 
\vspace{-10pt}
\subsection{Experimental Setup}
\textbf{Dataset and Evaluation Protocol.} We conduct our experiments with the MSD and follow the same data preprocessing procedure as in \cite{multivae} which only keeps the users who at least listened to 20 songs and the songs that are listened to by at least 200 users. Before this filtering stage, we also removed the songs from the dataset for which we do not have the audio files. Our dataset consists of 40,940 songs, 469,432 train users, 50,000 validation users and 50,000 test users. We conduct our evaluation in terms of strong generalization in which training, validation and test sets have disjoint users. We report the Normalized Discounted Cumulative Gain (\textit{NDCG@100}) as well as Recall (\textit{Recall@20}, \textit{Recall@50}) as they are  the standard performance metrics in the recommendation literature.

\noindent \textbf{Tags and Prototype Generation.} We select prototypes to correspond to the 80 most commonly used song-level tags according to the Last.fm Dataset \cite{bertin2011million}. Tags fall into 4 major groups: \textit{era}, \textit{genre}, \textit{mood}, and \textit{instrumentation}. We select the most listened songs in the dataset for each tag.

\noindent \textbf{Music Feature Extractor.} We extract music features for each song in the dataset and prototype songs with the MERT-v1-330M model \cite{mert}. We use the last representation layer (1024-$D$) in our experiments, which is found to give the best performance in terms of representation performance.

\noindent \textbf{Baselines.} APRON can be labeled as an autoencoder based method therefore we compare it with other autoencoder baselines. As baselines, we use MultiDAE, MultiVAE \cite{multivae}, RecVAE \cite{recvae} and
MacridVAE, SEM-MacridVAE \cite{macridvae} with our data split. We could not directly use the numbers from the corresponding papers as the version of the dataset does not contain the audio files for 200 songs audio files, and we have therefore run the baselines ourselves using the official repositories. 

\noindent \textbf{Implementation Details.} In our experiments, when implementing the attention mechanism to express each song in terms of prototypes in Eq. \ref{eq:attention}, we use multihead-attention. This results in the following way of calculating the protoype weights for each song: 
\begin{align}
    w_{kj,h}^i = \frac{\exp \left ( (P_{k,h} A_{k,h}^p)^\top (x_j^i A_{j,h}^x) \right )}{ \sum_{k'}^K \exp \left( (P_{k',h} A_{k',h}^p)^\top (x_j^i A_{j,h}^x) \right )} \label{eq:attention_mh},     
\end{align}
where we learn a matrix $A_{k,h}^p \in \mathbb R^{(D/nh)\times D'}$, $A_j^x \in \mathbb R^{(D/nh) \times D'}$, for each head $h$. The user profile is then calculated as, 
\begin{align}
    u_{i,h} &= \sum_{k=1}^K \underbrace{\sum_{j=1}^{S_i} \frac{\exp \left ( (P_{k,h} A_{k,h}^p)^\top (x_j^i A_{j,h}^x) \right )}{ \sum_{k'}^K \exp \left( (P_{k',h} A_{k',h}^p)^\top (x_j^i A_{j,h}^x) \right )} }_{:=\widehat w^i_{k,h}} \underbrace{(P_{k,h} A_{k,h}^p)}_{:=\widetilde P_{k,h}} \label{eq:user_mh} 
\end{align}
Note that $P_{k,h}$, is obtained by dividing the prototype vector into $H$ equal length chunks. Then to obtain the final user profile $u_i$, we concatenate over the head dimension $h$, such that,
\begin{align}
    u_i = \text{Concatenate}([u_{i, 1}, u_{i,2}, \dots, u_{i, H}]),
\end{align}
where $H$ is the number of attention heads.

\noindent \textbf{Controllability Metrics.} 
Besides the recommendation system performance, we also define a controllability metric based on NDCG as defined follows. For a specific tag $\tau$, we define the tag-wise $\operatorname{DCG}_t@k$ as follows (the subscript $t$ is used to denote tag-wise DCG), 
\begin{align}
    \operatorname{DCG}_t @ k(\tau)=\sum_{i=1}^k \frac{\mathbf{I}(\tau \in T(y_i))}{\log _2(i+1)},
\end{align}
 where $T(.)$ extracts the tag information that corresponds to song song $y_i$, and $\mathbf I(.)$ denotes the indicator function. That is, if the tag $\tau$ is contained in the tags of the song $y_i$ (denoted with $T(y_i)$, the indicator function $\mathbf I(.)$ returns 1).

Then we calculate $\operatorname{NDCG}$ for all users in $U_{\tau}$, where $U_{\tau}$ denotes the set of users having items with tag $\tau$:
\begin{align}
    \operatorname{NDCG}_t @ k(\tau) = \frac{1}{|U_{\tau}|}\sum_{u \in U_{\tau}}\sum_{i=1}^k \frac{\mathbf{I}(\tau \in T(i))}{\log _2(i+1)}. 
\end{align}

We define the controllability metric ($\Delta@k(\tau)$) to measure the interpretability performance of our system.
We calculate the change ($\Delta @ k$) between the full (using all of the templates, denoted with a superscript $F$) and modified (we denote with a superscript $M$). 
\begin{align}
\Delta@k(\tau) = \operatorname{NDCG}_t^{F} @ k(\tau) - \operatorname{NDCG}_t^{M} @ k(\tau) \label{eq:labeldelta-ndcg}
\end{align}
When we drop attention weights, we allow using all of the prototypes except the prototype that corresponds to the tag $\tau$.

In this calculation we filter the users with having 0 in both term since every user does not contribute to each tag. Results, averaged over all tags, are presented in Table \ref{tab-results}. We furthermore provide an analysis to breakdown the contribution of each tag in Figure \ref{fig:tagcontrol}.
\vspace{-5pt}
\subsection{Recommendation Performance}
\label{sec:controllability}
In Table \ref{tab-results}, we compare the recommendation performance of APRON and several baselines introduced in the previous section. We evaluate recommendation performance under strong-generalization (i.e for users not seen during training). We observe that APRON with an attention mechanism with 16 parallel head ($H=16$) is able to obtain competitive results in terms of NDCG. We use $\lambda_1$ as 1 and $\lambda_2$ as 0.005.

\setlength\heavyrulewidth{0.20ex}
\begin{table*}[ht]
\caption{Comparison of recommendation and controllability performance of APRON and baselines. Note that MacridVAE and SEM-MacridVAE are exact same methods except the initialization of item and prototype embeddings.}
\label{tab-results}
\begin{center}
  \begin{tabular}{lccccc}
    \toprule
Method   & Recall@20 ($\uparrow$) & Recall@50 ($\uparrow$) & NDCG@100 ($\uparrow$) & ${\Delta@20}$ ($\uparrow$) & ${\% \Delta@20}$ ($\uparrow$) \\ \midrule

$\text{MultiDAE} $\cite{multivae}  & 0.253 & 0.355 & 0.300 & N/A & N/A\\

$\text{MultiVAE}$ \cite{multivae}  & 0.264 & 0.366 & 0.315  & N/A & N/A \\ 

$\text{RecVAE} $ \cite{recvae}  & 0.275 & 0.373 & 0.325 & N/A & N/A \\
$\text{MacridVAE}$ \cite{macridvae}  & 0.291  & 0.385 & 0.343 & N/A & N/A \\
$\text{SEM-MacridVAE}$ \cite{macridvae}  & 0.290  & 0.383 & 0.341 & -0.00015 & -0.05   \\ 
\midrule

APRON (Ours)   & 0.277 &  0.377 & 0.327 & 0.05407 & 33.80 \\ \bottomrule  
\end{tabular}
\end{center}
\end{table*}
\vspace{-5pt}
\subsection{Controllability}
To assess the controllability of APRON, we conduct an experiment where we manipulate the attention weights $\widehat w^i_k$ for $k\in \{1, \dots, K\}$ that correspond to the different musical tags. The expectation is that if, for instance, the weight corresponding to tag $k$ is lowered,  songs associated with this tag would be less likely to be recommended. 

\begin{figure*}[ht]
    \centering
    \includegraphics[width=\linewidth]{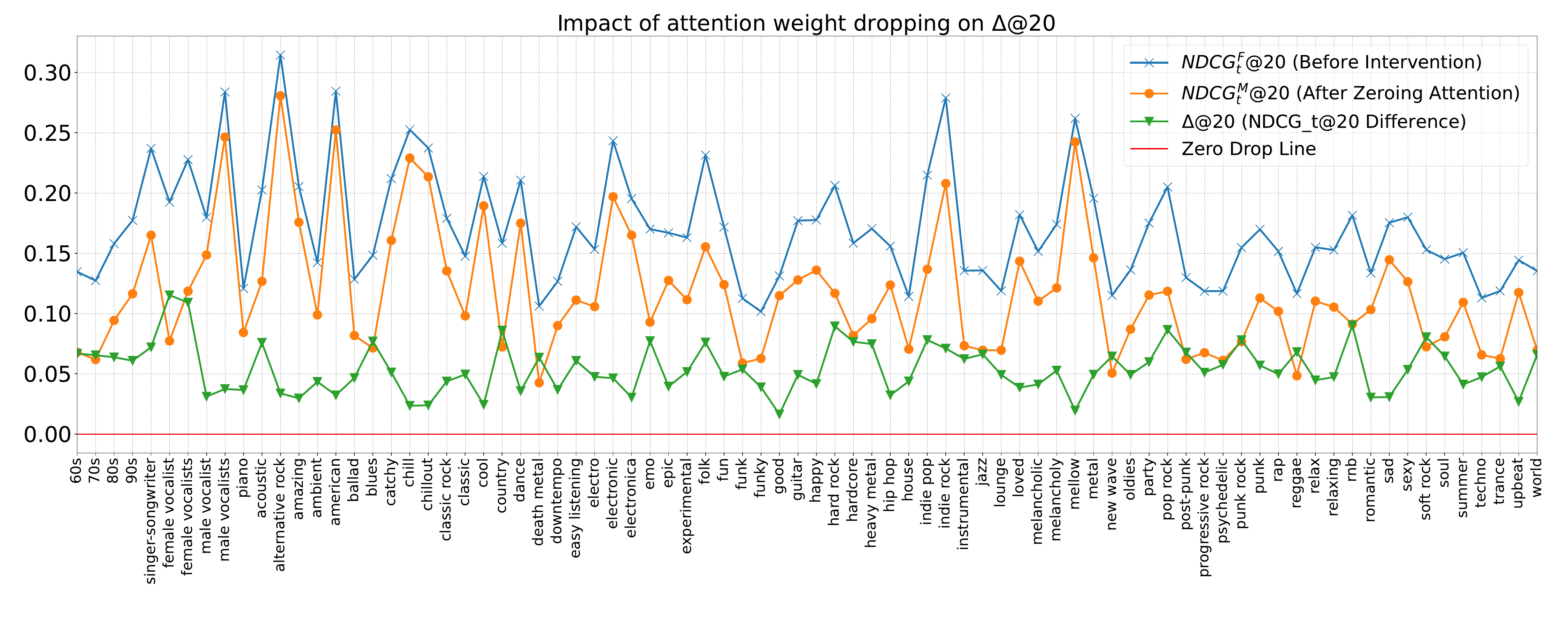}
    \vspace{-.7cm}
    \caption{Tag based controllability of APRON (for attention weight reduction): We observe that for almost all tags reducing the associated attention weight $\widehat w^i_k$ results in reduction in $NDCG_t$. This indicates that the recommendations are guided through prototypes. This experiment is conducted for $H=16$. }
    \label{fig:tagcontrol}
\end{figure*}

We showcase this in Figure \ref{fig:tagcontrol} where we systematically lower the weight associated with a tag and evaluate its effect on the recommendation quality. We observe that for almost all tags, reducing the attention weight to zero $\widetilde w^i_k$ for the tag $k$ results in a drop in $NDCG_t$ for that particular tag (using the metric defined in Eq.~\ref{eq:labeldelta-ndcg}). As reported in the last two columns of Table \ref{tab-results}, SEM-MacridVAE is the only comparable method among the baselines and achieves $\sim 0$ controllability, while APRON offers a decent level of controllability.

\section{Conclusions}
We have proposed APRON, a prototypical network for music recommendations. Experiments on the MSD show that APRON can produce controllable recommendations (more controllable compared to SEM-Macrid VAE, for example) while maintaining competitive recommendation performance with other baselines. All in all, APRON is a new form of scrutable recommendation system which directly exposes user modelling, paving the way for domain specific scrutable models that captures the feature level information.  
As future work, we would like to also apply APRON on other application domains where prototypes can be used to encode item characteristic difficult to encode using text (e.g.\ fashion recommendation). 


\bibliographystyle{IEEEbib}

\small
\bibliography{refs}

\end{document}